\documentclass[aps,rpl,preprint,groupedaddress]{revtex4}
\begin{document}

\title{Teleportation with a Mixed State of Four Qubits and the Generalized Singlet Fraction}
\author{Ye Yeo}
\affiliation{Department of Physics, National University of Singapore, 10 Kent Ridge Crescent, Singapore 119260, Singapore}

\begin{abstract}
Recently, an explicit protocol ${\cal E}_0$ for faithfully teleporting arbitrary two-qubit states using genuine four-qubit entangled states was presented by us [Phys. Rev. Lett. {\bf 96}, 060502 (2006)].  Here, we show that ${\cal E}_0$ with an arbitrary four-qubit mixed state resource $\Xi$ is equivalent to a generalized depolarizing bichannel with probabilities given by the maximally entangled components of the resource.  These are defined in terms of our four-qubit entangled states.  We define the generalized singlet fraction ${\cal G}[\Xi]$, and illustrate its physical significance with several examples.  We argue that in order to teleport arbitrary two-qubit states with average fidelity better than is classically possible, we have to demand that ${\cal G}[\Xi] > 1/2$.  In addition, we conjecture that when ${\cal G}[\Xi] < 1/4$ then no entanglement can be teleported.  It is shown that to determine the usefulness of $\Xi$ for ${\cal E}_0$, it is necessary to analyze ${\cal G}[\Xi]$.
\end{abstract}

\maketitle

Many of the profound results in quantum information theory \cite{Nielsen} are impossible without the resource of quantum entanglement.  For instance, it enables one to implement all possible global quantum operations locally by making use of the concept of quantum teleportation - the subject of this paper (see, for example, Ref.\cite{Huang}).  To fully utilize this resource, which may be found in natural systems in numerous different forms, we must first understand, in general, the various different kinds of multipartite entanglement that allow specific quantum information processing tasks to be successfully carried out.

Bennett {\em et al.} \cite{Bennett1} are the first to show how bipartite quantum entanglement can assist in the teleportation of an intact quantum state $|\psi\rangle_{A_1} = a|0\rangle_{A_1} + b|1\rangle_{A_1}$, with $a, b \in {\cal C}^1$ and $|a|^2 + |b|^2 = 1$, from one place to another, by a sender, Alice, who knows neither the state $|\psi\rangle_{A_1}$ to be teleported nor the location of the intended receiver, Bob.  In their standard teleportation protocol ${\cal T}_0$, Alice and Bob share {\em a priori} a pair of particles, $A_2$ and $B$, in a maximally entangled Bell state, say $|\Psi^0_{\rm Bell}\rangle_{A_2B} \equiv (|00\rangle_{A_2B} + |11\rangle_{A_2B})/\sqrt{2}$.  Perfect teleportation is possible only when maximally entangled pure channel states are available.  However, due to the undesired coupling of the quantum states with the environment, we have to deal with entangled mixed channel states in practical situations.  ${\cal T}_0$, when used with an arbitrary two-qubit mixed state $\chi_{A_2B}$ as a resource, acts as a generalized depolarizing channel \cite{Bowen, Albeverio},
\begin{equation}
\Lambda^{\chi, {\cal T}_0}_B(|\psi\rangle_B\langle\psi|) = \sum^3_{\mu = 0}
\langle\Psi^{\mu}_{\rm Bell}|\chi|\Psi^{\mu}_{\rm Bell}\rangle \times u^{\mu\dagger}_B|\psi\rangle_B\langle\psi|u^{\mu}_B,
\end{equation}
where $|\Psi^{\mu}_{\rm Bell}\rangle_{AB} = (u^{\mu}_A \otimes u^0_B)|\Psi^0_{\rm Bell}\rangle_{AB}$ $(\mu = 0, 1, 2, 3)$; $u^0$ is the two-dimensional identity, $u^1 = \sigma^1$, $u^2 = i\sigma^2$, and $u^3 = \sigma^3$.  Here, $\sigma^j$ $(j = 1, 2, 3)$ are the Pauli matrices.  It will be useful to note that $|ij\rangle_{AB} = \frac{1}{\sqrt{2}}\sum^3_{\mu = 0}(u^{\mu\dagger})_{ji}|\Psi^{\mu}_{\rm Bell}\rangle_{AB}$, since $\sum^3_{\mu = 0}(u^{\mu\dagger})_{ji}(u^{\mu})_{mn} = 2\delta_{im}\delta_{jn}$.  A slightly more general teleportation protocol ${\cal T}_1$ yields \cite{Albeverio}
\begin{equation}
\Lambda^{\chi, {\cal T}_1}_B(|\psi\rangle_B\langle\psi|) = \frac{1}{4}\sum^3_{\alpha, \beta = 0}
\langle\Psi^{\alpha}_{\rm Bell}|\chi|\Psi^{\beta}_{\rm Bell}\rangle
\sum^3_{\mu = 0} r^{\mu}_Bu^{\alpha\dagger}_Bu^{\mu\dagger}_B
|\psi\rangle_B\langle\psi|u^{\mu}_Bu^{\beta}_Br^{\mu\dagger}.
\end{equation}
Note that Eq.(2) reduces to Eq.(1) when Bob chooses unitary $r^{\mu} = u^{\mu}$, and ${\cal T}_1$ becomes ${\cal T}_0$.  The fidelity of teleportation \cite{Albeverio}
\begin{eqnarray}
\Phi[\Lambda^{\chi, {\cal T}_1}_B] & \equiv & \int d\psi\ 
{_{B}}\langle\psi|\Lambda^{\chi, {\cal T}_1}_B(|\psi\rangle_B\langle\psi|)|\psi\rangle_B \nonumber \\
& = & \frac{1}{3} + \frac{1}{6}\sum^3_{\mu = 0}\langle\Psi^0_{\rm Bell}|(u^0 \otimes u^{\mu\dagger}r^{\mu})\chi
(u^0 \otimes r^{\mu\dagger}u^{\mu})|\Psi^0_{\rm Bell}\rangle.
\end{eqnarray}
Clearly, $\Phi[\Lambda^{\chi, {\cal T}_0}_B] = 1/3 + 2{\cal F}[\chi]/3$, where the singlet fraction 
\begin{equation}
{\cal F}[\chi] \equiv \langle\Psi^0_{\rm Bell}|\chi|\Psi^0_{\rm Bell}\rangle.
\end{equation}
Furthermore, since the group of unitary transformations in finite dimensions is compact, there exists $r^{\mu} = r^{\mu}_{\rm opt}$ such that we have the maximal singlet fraction 
\begin{eqnarray}
{\cal F}_{\max}[\chi] & \equiv & \max_u
\langle\Psi^0_{\rm Bell}|(u^0 \otimes u)\chi(u^0 \otimes u^{\dagger})|\Psi^0_{\rm Bell}\rangle \nonumber \\
& = & \langle\Psi^0_{\rm Bell}|(u^0 \otimes u^{\mu\dagger}r^{\mu}_{\rm opt})\chi
(u^0 \otimes r^{\mu\dagger}_{\rm opt}u^{\mu})|\Psi^0_{\rm Bell}\rangle;
\end{eqnarray}
and hence the maximal teleportation fidelity $\Phi[\Lambda^{\chi, {\cal T}_{\rm opt}}_B] =  1/3 + 2{\cal F}_{\max}[\chi]/3$.  The maximization in Eq.(5) is over the set of all unitary operations $u$ on ${\cal C}^2$.  Teleportation thus firmly establishes the practical basis for considering the maximally entangled Bell states as basic units, upon which bipartite entanglement can be quantitatively expressed in terms of.  In particular, the singlet fraction ${\cal F}[\chi]$ determines quantitatively the suitability of a given two-qubit mixed state $\chi$ as a resource for ${\cal T}_0$ \cite{Horodecki}.

The teleportation of an arbitrary two-qubit state, $|\Psi\rangle_{A_1A_2} = \sum^1_{i, j = 0}a_{ij}|ij\rangle_{A_1A_2}$, with $a_{ij}\in {\cal C}^1$ and $\sum^1_{i, j = 0}|a_{ij}|^2 = 1$, had been studied by Lee {\em et al.} \cite{Lee} and by Rigolin \cite{Rigolin}.  Recently, we gave, in Ref.\cite{Yeo}, an explicit protocol ${\cal E}_0$ for faithfully teleporting arbitrary two-qubit states employing genuine four-qubit entangled states
\begin{equation}
|\Upsilon^{00}(\theta_{12}, \phi_{12})\rangle_{A_3A_4B_1B_2} \equiv 
\frac{1}{2}\sum^3_{J = 0}|J\rangle_{A_3A_4} \otimes |J'\rangle_{B_1B_2}.
\end{equation}
The $|J\rangle$'s constitute an orthonormal basis, and explicitly $|J\rangle = S|ij\rangle$ with
\begin{equation}
S(\theta_1, \phi_1) \equiv \left(\begin{array}{cccc}
\cos\theta_1 & 0 & 0 & -\sin\theta_1 \\
0 & \cos\phi_1 & -\sin\phi_1 & 0 \\
0 & \sin\phi_1 & \cos\phi_1 & 0 \\
\sin\theta_1 & 0 & 0 & \cos\theta_1
\end{array}\right).
\end{equation}
The $|J'\rangle$'s constitute another orthonormal basis: $|J'\rangle = T|ij\rangle$ with
\begin{equation}
T(\theta_2, \phi_2) \equiv \left(\begin{array}{cccc}
\cos\theta_2 & 0 & 0 & -\sin\theta_2 \\
0 & \sin\phi_2 & \cos\phi_2 & 0 \\
0 & \cos\phi_2 & -\sin\phi_2 & 0 \\
\sin\theta_2 & 0 & 0 & \cos\theta_2
\end{array}\right).
\end{equation}
In other word,
\begin{equation}
|\Upsilon^{00}(\theta_{12}, \phi_{12})\rangle_{A_3A_4B_1B_2} = \frac{1}{\sqrt{2}}
(|\zeta^0(\theta_{12}, \phi_{12})\rangle + |\zeta^1(\theta_{12}, \phi_{12})\rangle)_{A_3A_4B_1B_2},
\end{equation}
with $|\zeta^0\rangle \equiv (\cos\theta_{12}|0000\rangle - \sin\theta_{12}|0011\rangle - \sin\phi_{12}|0101\rangle + \cos\phi_{12}|0110\rangle)/\sqrt{2}$ and $|\zeta^1\rangle \equiv (\cos\phi_{12}|1001\rangle + \sin\phi_{12}|1010\rangle + \sin\theta_{12}|1100\rangle + \cos\theta_{12}|1111\rangle)/\sqrt{2}$.  Here, $-\pi/2 < \theta_{12} \equiv \theta_1 - \theta_2 < \pi/2$ and $-\pi/2 < \phi_{12} \equiv \phi_1 - \phi_2 < \pi/2$, since $0 < \theta_1,\ \theta_2,\ \phi_1,\ \phi_2 < \pi/2$.  $|\Upsilon^{00}\rangle$ is a genuine multipartite entangled state in the sense of \cite{Osterloh}.  In fact, the third-, fourth- and sixth-order four-qubit filters: ${\cal F}^{(4)}_1$, ${\cal F}^{(4)}_2$ and ${\cal F}^{(4)}_3$ have the following respective expectation values for $|\Upsilon^{00}\rangle$:
\begin{eqnarray}
\langle\Upsilon^{00}|{\cal F}^{(4)}_1|\Upsilon^{00}\rangle & \equiv & 
\sum^3_{\alpha,\beta,\gamma = 0}\delta_{\alpha_1\beta_1}\delta_{\alpha_2\gamma_1}\delta_{\beta_2\gamma_2}
E_{\alpha_1\alpha_2}E^{\beta_1}_{\ \beta_2}E^{\gamma_1\gamma_2} \nonumber \\
& = & \frac{1}{2}(\cos 2\theta_{12}\sin^22\phi_{12} + \cos 2\phi_{12}\sin^22\theta_{12}), \nonumber \\
\langle\Upsilon^{00}|{\cal F}^{(4)}_2|\Upsilon^{00}\rangle & \equiv & \sum^3_{\alpha,\beta,\delta, \epsilon = 0}
\delta_{\alpha_1\beta_1}\delta_{\alpha_2\delta_1}\delta_{\beta_2\epsilon_1}\delta_{\delta_2\epsilon_2}
E_{\alpha_1\alpha_2}E^{\beta_1}_{\ \beta_2}E^{\delta_1}_{\ \delta_2}E^{\epsilon_1\epsilon_2} \nonumber \\
& = & \frac{1}{2}\sin^22\theta_{12}\sin^22\phi_{12} + \frac{1}{4}(1 - \cos 2\theta_{12}\cos 2\phi_{12})(\sin^22\theta_{12} + \sin^22\phi_{12}), \nonumber \\
\langle\Upsilon^{00}|{\cal F}^{(4)}_3|\Upsilon^{00}\rangle & \equiv & \frac{1}{2}\sum^3_{\alpha,\beta,\gamma = 0}
E^{\alpha_1\alpha_2}E_{\alpha_1\alpha_2}E^{\beta_1\beta_2}E_{\beta_1\beta_2}E^{\gamma_1\gamma_2}E_{\gamma_1\gamma_2} \nonumber \\
& = & \frac{1}{8}(1 - 2\cos 2\theta_{12}\cos 2\phi_{12})(2 + \cos 2\theta_{12}\cos 2\phi_{12}) \nonumber \\
& & \times [2(\sin^22\theta_{12} + \sin^22\phi_{12}) - (\cos 2\theta_{12} - \cos 2\phi_{12})^2],
\end{eqnarray}
which become identically zero only when $\theta_{12} = \phi_{12} = 0$.  Here, 
$E^{\alpha_1\alpha_2} \equiv 
\langle\Upsilon^{00}|\sigma^{\alpha_1}\otimes\sigma^{\alpha_2}\otimes\sigma^2\otimes\sigma^2|\Upsilon^{00}\rangle$, 
$E^{\beta_1\beta_2} \equiv 
\langle\Upsilon^{00}|\sigma^{\beta_1}\otimes\sigma^2\otimes\sigma^{\beta_2}\otimes\sigma^2|\Upsilon^{00}\rangle$, 
$E^{\gamma_1\gamma_2} \equiv 
\langle\Upsilon^{00}|\sigma^2\otimes\sigma^{\gamma_1}\otimes\sigma^{\gamma_2}\otimes\sigma^2|\Upsilon^{00}\rangle$, 
$E^{\delta_1\delta_2} \equiv 
\langle\Upsilon^{00}|\sigma^2\otimes\sigma^{\delta_1}\otimes\sigma^2\otimes\sigma^{\delta_2}|\Upsilon^{00}\rangle$, 
$E^{\epsilon_1\epsilon_2} \equiv 
\langle\Upsilon^{00}|\sigma^2\otimes\sigma^2\otimes\sigma^{\epsilon_1}\otimes\sigma^{\epsilon_2}|\Upsilon^{00}\rangle$; 
and $E_{\kappa\lambda} = g_{\kappa\mu}g_{\lambda\nu}E^{\mu\nu}$ with $g_{\mu\nu} \equiv {\rm diag}\{-1,\ 1,\ 0,\ 1\}$.  The expectation values of ${\cal F}^{(4)}_1$, ${\cal F}^{(4)}_2$ and ${\cal F}^{(4)}_3$ are $1$, $1$ and $1/2$ respectively for the four-qubit GHZ state \cite{Greenberger}; but are identically zero for the four-qubit W state \cite{Zeilinger}.  Therefore, $|\Upsilon^{00}\rangle$, W and GHZ states are inequivalent under stochastic local operations and classical communication (SLOCC).  Both the GHZ and W states do not enable the faithful teleportation of an arbitrary two-qubit state.  Whenever $\theta_{12} = \phi_{12} = 0$, $|\Upsilon^{00}\rangle$ is reducible to a tensor product of two Bell states: $|\Upsilon^{00}\rangle_{A_3A_4B_1B_2} = |\Psi^0_{\rm Bell}\rangle_{A_3B_2} \otimes |\Psi^0_{\rm Bell}\rangle_{A_4B_1}$, which, with $B_1$ and $B_2$ interchanged, gives $|g_1\rangle_{A_3A_4B_1B_2}$ in \cite{Rigolin}.  As mentioned above, the expectation values of ${\cal F}^{(4)}_1$, ${\cal F}^{(4)}_2$ and ${\cal F}^{(4)}_3$ are identically zero in this case.  Consequently, $|\Upsilon^{00}\rangle$ is not SLOCC equivalent to a tensor product of two Bell states, even though it also enables the faithful teleportation of an arbitrary two-qubit state.  Here, we define the generalized Smolin states
\begin{equation}
\Xi^{\rm GS}(\theta_{12}, \phi_{12}) \equiv \frac{1}{4}\sum^3_{\mu = 0}
(U^{00} \otimes U^{\mu\mu\dagger})
|\Upsilon^{00}(\theta_{12}, \phi_{12})\rangle\langle\Upsilon^{00}(\theta_{12}, \phi_{12})|(U^{00} \otimes U^{\mu\mu}),
\end{equation}
where $U^{\mu\nu} \equiv u^{\mu} \otimes u^{\nu}$.  $\Xi^{\rm GS}(\theta_{12}, \phi_{12})$ reduces to the Smolin state $\Xi^{\rm S}$ \cite{Smolin} when $\theta_{12} = \phi_{12} = 0$.  In ${\cal E}_0$, Alice performs a complete projective measurement jointly on $A_1A_2A_3A_4$ in the following basis of 16 orthonormal states \cite{Yeo}:
\begin{eqnarray}
|\Pi^{\mu\nu}(\theta_{12}, \phi_{12})\rangle_{A_1A_2A_3A_4} 
& \equiv & (U^{\mu\nu}_{A_1A_2} \otimes U^{00}_{A_3A_4})|\Pi^{00}(\theta_{12}, \phi_{12})\rangle_{A_1A_2A_3A_4} \nonumber \\
& = & \frac{1}{2}\sum^1_{i, j, k, l = 0} (u^{\mu})_{ik}(u^{\nu})_{jl}
U^{\mu\nu}_{A_1A_2}TU^{\mu\nu\dagger}_{A_1A_2}|ij\rangle_{A_1A_2} \otimes S|kl\rangle_{A_3A_4},
\end{eqnarray}
since $|\Pi^{00}(\theta_{12}, \phi_{12})\rangle_{A_1A_2A_3A_4} \equiv \frac{1}{2}\sum^3_{K = 0}|K'\rangle_{A_1A_2} \otimes |K\rangle_{A_3A_4}$.  It follows that
\begin{equation}
|ijkl\rangle_{A_1A_2A_3A_4} = \frac{1}{2}\sum^3_{\mu, \nu = 0}(u^{\mu\dagger})_{ki} (u^{\nu\dagger})_{lj}
[(U^{\mu\nu}_{A_1A_2}T^{-1} \otimes S^{-1})|\Pi^{00}(\theta_{12}, \phi_{12})\rangle_{A_1A_2A_3A_4}].
\end{equation}
Upon receiving four bits of classical information about Alice's measurement result, Bob can always succeed in recovering an exact replica of the original state of her particles $A_1A_2$ by applying the appropriate ``recovery'' unitary operations on his particles $B_1B_2$.

In this paper, we derive the equation [Eq.(28)] that corresponds to Eq.(2) when Alice and Bob share {\em a priori} two pairs of particles, $A_3A_4$ and $B_1B_2$, in an arbitrary four-qubit mixed state, $\chi_{A_3A_4B_1B_2}$, as a resource.  For our teleportation protocol ${\cal E}_0$, it reduces to
\begin{equation}
\Lambda^{\Xi, {{\cal E}_0}}_{B_1B_2}(|\Psi\rangle_{B_1B_2}\langle\Psi|) = 
\sum^3_{\mu, \nu = 0}\langle\Upsilon^{\mu\nu}|\Xi|\Upsilon^{\mu\nu}\rangle 
\times U^{\mu\nu\dagger}_{B_1B_2}|\Psi\rangle_{B_1B_2}\langle\Psi|U^{\mu\nu}_{B_1B_2},
\end{equation}
where, analogous to $|\Pi^{\mu\nu}\rangle$, we define $|\Upsilon^{\mu\nu}\rangle \equiv (U^{00} \otimes U^{\mu\nu\dagger})|\Upsilon^{00}\rangle$.  That is, ${\cal E}_0$ with $\Xi_{A_3A_4B_1B_2}$, acts as a generalized depolarizing bichannel.  In addition, we derive the fidelity of teleportation
\begin{eqnarray}
\Phi[\Lambda^{\Xi, {\cal E}_1}_{B_1B_2}] & \equiv & \int d\Psi\ 
{_{B_1B_2}}\langle\Psi|\Lambda^{\Xi, {\cal E}_1}_{B_1B_2}(|\Psi\rangle_{B_1B_2}\langle\Psi|)|\Psi\rangle_{B_1B_2} \nonumber \\
& = & \frac{1}{5} + \frac{1}{20}\sum^3_{\mu, \nu = 0}\langle\Upsilon^{00}|
(U^{00} \otimes U^{\mu\nu\dagger}R^{\mu\nu})\Xi(U^{00} \otimes R^{\mu\nu\dagger}U^{\mu\nu})|\Upsilon^{00}\rangle.
\end{eqnarray}
In ${\cal E}_0$, we have $R^{\mu\nu} = U^{\mu\nu}$ and hence $\Phi[\Lambda^{\Xi, {\cal E}_0}_{B_1B_2}] = 1/5 + 4\langle\Upsilon^{00}|\Xi|\Upsilon^{00}\rangle/5$.  For it to be a useful quantity, we define the generalized singlet fraction
\begin{equation}
{\cal G}[\Xi] \equiv \max_{\theta_{12}, \phi_{12}}
\{\langle\Upsilon^{00}(\theta_{12}, \phi_{12})|\Xi|\Upsilon^{00}(\theta_{12}, \phi_{12})\rangle\},
\end{equation}
in contrast to Eq.(4).  The $\theta_{12}$ and $\phi_{12}$ that give ${\cal G}$, determine Alice's appropriate measurement, Eq.(12).  In Ref.\cite{Horodecki} Horodecki {\em et al.} have shown that, for a given $N \times N$ channel state $\chi$, singlet fraction ${\cal F}[\chi] < 1/N$ implies that one cannot do standard teleportation ${\cal T}_0$ with $\chi$ with better than classical fidelity $\Phi_{\rm class} = 2/(N + 1)$.  This is the maximum value of the mean fidelity of the estimated state vector in an $N$-dimensional Hilbert space \cite{Derka}.  For $N = 2$, $\Phi_{\rm class} = 2/3$.  It is also the maximum possible value of the fidelity of output qubits from optimal quantum cloning machines \cite{Gisin}.  In Ref.\cite{Buzek}, it is shown that the maximum possible value of the fidelity of the output states of two qubits from optimal nonlocal quantum cloning machines is $3/5$.  For the teleportation of two-qubit states, we thus consider $\Phi_{\rm class} = 3/5$ and hence ${\cal G}_{\rm crit} = 1/2$.  Therefore, in assessing the suitability of a given entangled mixed state $\Xi$ of four qubits for ${\cal E}_0$, it is necessary to calculate ${\cal G}[\Xi]$.  We shall illustrate the physical significance of the generalized singlet fraction with a few examples before giving the derivations and further discussions.

As a first example, let us consider the mixed state
\begin{equation}
\Xi(\alpha, \beta) = q|\Upsilon^{00}(\alpha, \beta)\rangle\langle\Upsilon^{00}(\alpha, \beta)| + \frac{1 - q}{16}I,
\end{equation}
where $0 \leq q \leq 1$ and $I$ is the sixteen-dimensional identity.  We have
\begin{equation}
\langle\Upsilon^{00}(\theta_{12}, \phi_{12})|\Xi(\alpha, \beta)|\Upsilon^{00}(\theta_{12}, \phi_{12})\rangle = 
\frac{1 - q}{16} + \frac{q}{4}[\cos(\theta_{12} - \alpha) + \cos(\phi_{12} - \beta)]^2,
\end{equation}
which has the maximal value ${\cal G}[\Xi] = (1 + 15q)/16$ when $\theta_{12} = \alpha$ and $\phi_{12} = \beta$.  Specifically, if $\alpha = \beta = 0$, then $|\Upsilon^{00}(\alpha, \beta)\rangle$ is essentially a tensor product of two EPR channels: $|\Psi^0_{\rm Bell}\rangle \otimes |\Psi^0_{\rm Bell}\rangle$, and Alice performs a measurement in the basis made up of tensor products of Bell states.  Otherwise, to achieve ${\cal G}[\Xi]$, she will have to perform a measurement in the basis $\{|\Pi^{\mu\nu}(\alpha, \beta)\rangle\}$.  Clearly, ${\cal G}[\Xi] = 1/2$ when $q_{\rm crit} = 7/15$.  Suppose $|\Psi\rangle_{A_1A_2} = \cos\epsilon|00\rangle_{A_1A_2} + \sin\epsilon|11\rangle_{A_1A_2}$ with $0 \leq \epsilon \leq \pi/2$, the negativity \cite{Vidal} of the teleported state
\begin{equation}
{\cal N}[\Lambda^{\Xi, {\cal E}_0}_{B_1B_2}(|\Psi\rangle_{B_1B_2}\langle\Psi|)] = \max\{0,\ -\frac{1}{2}(1 - q) + q\sin2\epsilon\},
\end{equation}
which is zero whenever $q \leq 1/5 < q_{\rm crit}$ or ${\cal G}[\Xi] = 1/4$.  It indicates that when the teleportation fidelity $\Phi[\Lambda^{\Xi, {\cal E}_0}_{B_1B_2}] \leq 2/5 < \Phi_{\rm class}$, then no entanglement may be teleported.  If $\epsilon = \pi/12$, then ${\cal N}[\Lambda^{\chi, {\cal E}_0}_{B_1B_2}(|\Psi\rangle_{B_1B_2}\langle\Psi|)] = 0$ when $q \leq 1/2$.  We thus conclude that even when we have nonclassical teleportation fidelity, the entanglement of two-qubit states with entanglement smaller than some critical amount, say ${\cal N}_{\rm crit}$, may become zero in ${\cal E}_0$.  These states are being teleported  to separable states with average fidelities that are nevertheless not achievable by ``classical'' means.  Entanglement is fragile to teleport \cite{Lee2}.

Next, we consider the mixed state
\begin{equation}
\Xi(\alpha, \beta, \gamma, \delta) = 
q|\Upsilon^{00}(\alpha, \beta)\rangle\langle\Upsilon^{00}(\alpha, \beta)| + (1 - q)\Xi^{\rm GS}(\gamma, \delta).
\end{equation}
After some straightforward calculations as above, we obtain
\begin{eqnarray}
& & \langle\Upsilon^{00}(\theta_{12}, \phi_{12})|\Xi(\alpha, \beta, \gamma, \delta)|\Upsilon^{00}(\theta_{12}, \phi_{12})\rangle 
\nonumber \\
& = & \frac{q}{4}[\cos(\theta_{12} - \alpha) + \cos(\phi_{12} - \beta)]^2 
+ \frac{1 - q}{8}[\cos^2(\theta_{12} - \gamma) + \cos^2(\phi_{12} - \delta)].
\end{eqnarray}
For $q = 1$, we obviously have ${\cal G}[\Xi] = 1$ if $\theta_{12} = \alpha$ and $\phi_{12} = \beta$.  But, for $q = 0$, we have ${\cal G}[\Xi] = 1/4$ when $\theta_{12} = \gamma$ and $\phi_{12} = \delta$.  Therefore, ${\cal E}_0$ when used with a generalized Smolin state as a resource does not yield better than classical fidelity.  Clearly, to achieve ${\cal G}[\Xi]$ for $0 < q < 1$, depending on $\alpha$, $\beta$, $\gamma$ and $\delta$; $\theta_{12}$ and $\phi_{12}$ will in general be different functions of $q$.  This is in contrast to Eq.(18).  For definiteness, we study the case where $\alpha = \beta = \pi/4$ and $\gamma = \delta = 0$.  Then, if say $\theta_{12} = \phi_{12} = \arccos[(1 - q)/\sqrt{17q^2 - 2q + 1}]$, we have
\begin{equation}
{\cal G}[\Xi] = \frac{1}{8}(1 + 3q + \sqrt{17q^2 - 2q + 1}),
\end{equation}
which yields nonclassical teleportation fidelity for all possible values of $0.414214 = q_{\rm crit} < q < 1$.  For $|\Psi\rangle_{A_1A_2} = \cos\epsilon|00\rangle_{A_1A_2} + \sin\epsilon|11\rangle_{A_1A_2}$, the negativity \cite{Vidal} of the teleported state
\begin{equation}
{\cal N}[\Lambda^{\Xi, {\cal E}_0}_{B_1B_2}(|\Psi\rangle_{B_1B_2}\langle\Psi|)] = 
\max\{0,\ \frac{5q^2 - 2q + 1}{\sqrt{17q^2 - 2q + 1}}\sin2\epsilon\},
\end{equation}
which is nonzero for all $0 < q < 1$.  Hence, we conclude that, below $q_{\rm crit}$, the entanglement of some states may still be teleported even though they are done so with average fidelity below $\Phi_{\rm class}$.  

The above examples demonstrate the characteristic features of two-qubit teleportation.  These are quantitatively described by the generalized singlet fraction.  Namely, ${\cal G}[\Xi] > {\cal G}_{\rm crit} = 1/2$, which gives rise to nonclassical teleportation fidelity $\Phi[\Lambda^{\Xi, {\cal E}_0}_{B_1B_2}] > \Phi_{\rm class} = 3/5$; but the teleported states may or may not have entanglement.  And, $1/4 < {\cal G}[\Xi] \leq {\cal G}_{\rm crit}$, which yields teleportation fidelity $2/5 < \Phi[\Lambda^{\Xi, {\cal E}_0}_{B_1B_2}] \leq \Phi_{\rm class}$; but the teleported states may still have nonzero entanglement.  We conjecture that if ${\cal G}[\Xi] \leq 1/4$ and hence $\Phi[\Lambda^{\Xi, {\cal E}_0}_{B_1B_2}] < 2/5$, then no entanglement may be teleported.  The generalized singlet fraction can thus be used as a quantitative indicator of the usefulness of a given four-qubit entangled mixed state as a resource for ${\cal E}_0$.

Now we give the formal derivations before discussing further the physical significance of ${\cal G}[\Xi]$.  Suppose $\Xi_{A_3A_4B_1B_2} = \sum_{\lambda}p_{\lambda}|\xi^{(\lambda)}\rangle_{A_3A_4B_1B_2}\langle\xi^{(\lambda)}|$, where $0 \leq p_{\lambda} \leq 1$, $\sum_{\lambda}p_{\lambda} = 1$, and $|\xi^{(\lambda)}\rangle_{A_3A_4B_1B_2} = \sum^1_{k, l, m, n = 0}(C^{(\lambda)})_{mnkl}|kl\rangle_{A_3A_4} \otimes |mn\rangle_{B_1B_2}$.  Then, the initial complete state of the six particles, $A_1$, $A_2$, $A_3$, $A_4$, $B_1$ and $B_2$, is given by $|\Psi\rangle_{A_1A_2}\langle\Psi| \otimes \Xi_{A_3A_4B_1B_2} = \sum_{\lambda}p_{\lambda}(|\Psi\rangle_{A_1A_2} \otimes |\xi^{(\lambda)}\rangle_{A_3A_4B_1B_2})({_{A_1A_2}}\langle\Psi| \otimes {_{A_3A_4B_1B_2}}\langle\xi^{(\lambda)}|)$, with
\begin{equation}
|\Psi\rangle_{A_1A_2} \otimes |\xi^{(\lambda)}\rangle_{A_3A_4B_1B_2} = \frac{1}{2}\sum^3_{\mu, \nu = 0}
(U^{\mu\nu}_{A_1A_2}T^{-1} \otimes S^{-1})|\Pi^{00}\rangle_{A_1A_2A_3A_4} 
\otimes C^{(\lambda)}U^{\mu\nu\dagger}_{B_1B_2}|\Psi\rangle_{B_1B_2}.
\end{equation}
Eq.(24) follows from Eq.(13).  Straightforward calculations then yield
\begin{equation}
{_{A_1A_2A_3A_4}}\langle\Pi^{\mu\nu}|(|\Psi\rangle_{A_1A_2} \otimes |\xi^{(\lambda)}\rangle_{A_3A_4B_1B_2}) = 
\frac{1}{2}C^{(\lambda)}ST^{-1}U^{\mu\nu\dagger}_{B_1B_2}|\Psi\rangle_{B_1B_2}.
\end{equation}
Incidentally, if $|\xi^{(\lambda)}\rangle = |\Upsilon^{00}\rangle$, then $C^{(\lambda)} = TS^{-1}/2$; and hence ${_{A_1A_2A_3A_4}}\langle\Pi^{00}|(|\Psi\rangle_{A_1A_2} \otimes |\xi^{(\lambda)}\rangle_{A_3A_4B_1B_2}) = |\Psi\rangle_{B_1B_2}/4$ \cite{Yeo}.  Upon receiving 4 bits of classical information about Alice's measurement result $\mu\nu$, Bob applies the appropriate ``recovery'' unitary operations $R^{\mu\nu}_{B_1B_2}$ to his qubits $B_1B_2$ and hence obtains
\begin{equation}
\rho_{B_1B_2} = \frac{1}{4}\sum_{\lambda}p_{\lambda}\sum^3_{\mu, \nu = 0}
(R^{\mu\nu}_{B_1B_2}C^{(\lambda)}ST^{-1}U^{\mu\nu\dagger}_{B_1B_2})|\Psi\rangle_{B_1B_2}\langle\Psi|
(U^{\mu\nu}_{B_1B_2}TS^{-1}C^{(\lambda)\dagger}R^{\mu\nu\dagger}_{B_1B_2}).
\end{equation}
We recall that
\begin{equation}
\langle\Upsilon^{\alpha\beta}|\Xi|\Upsilon^{\gamma\delta}\rangle = \frac{1}{4}\sum_{\lambda} p_{\lambda}
{\rm tr}[U^{\alpha\beta\dagger}C^{(\lambda)}ST^{-1}]\ {\rm tr}[U^{\gamma\delta}TS^{-1}C^{(\lambda)\dagger}].
\end{equation}
Therefore, we may rewrite Eq.(26) as
\begin{equation}
\rho_{B_1B_2} = \frac{1}{16}\sum^3_{\alpha, \beta, \gamma, \delta = 0} 
\langle\Upsilon^{\alpha\beta}|\Xi|\Upsilon^{\gamma\delta}\rangle
\sum^3_{\mu, \nu = 0}
(R^{\mu\nu}_{B_1B_2}U^{\alpha\beta\dagger}_{B_1B_2}U^{\mu\nu\dagger}_{B_1B_2})
|\Psi\rangle_{B_1B_2}\langle\Psi|(U^{\mu\nu}_{B_1B_2}U^{\gamma\delta}R^{\mu\nu\dagger}_{B_1B_2}),
\end{equation}
which for $R^{\mu\nu} = U^{\mu\nu}$ clearly reduces to Eq.(14).  The teleportation fidelity
\begin{equation}
\Phi[\Lambda^{\Xi, {\cal E}_1}_{B_1B_2}] = \frac{1}{16}\sum^3_{\alpha, \beta, \gamma, \delta = 0} 
\langle\bar{\chi}^{\alpha\beta}|\chi|\bar{\chi}^{\gamma\delta}\rangle\sum^3_{\mu, \nu = 0}\int d\Psi\
\langle\Psi|R^{\mu\nu}U^{\alpha\beta\dagger}U^{\mu\nu\dagger}|\Psi\rangle
\langle\Psi|U^{\mu\nu}U^{\gamma\delta}R^{\mu\nu\dagger}|\Psi\rangle,
\end{equation}
with
\begin{eqnarray}
& & \int d\Psi\langle\Psi|R^{\mu\nu}U^{\alpha\beta\dagger}U^{\mu\nu\dagger}|\Psi\rangle
\langle\Psi|U^{\mu\nu}U^{\gamma\delta}R^{\mu\nu\dagger}|\Psi\rangle \nonumber \\
& = & \frac{1}{20}{\rm tr}[R^{\mu\nu}U^{\alpha\beta\dagger}U^{\mu\nu\dagger}]{\rm tr}[U^{\mu\nu}U^{\gamma\delta}R^{\mu\nu\dagger}]
+ \frac{1}{20}{\rm tr}[R^{\mu\nu}U^{\alpha\beta\dagger}U^{\mu\nu\dagger}U^{\mu\nu}U^{\gamma\delta}R^{\mu\nu\dagger}] \nonumber \\
& = & \frac{1}{5}\delta_{\alpha\gamma}\delta_{\beta\delta} + \frac{4}{5}
\langle\Upsilon^{00}|(U^{00} \otimes U^{\mu\nu\dagger}R^{\mu\nu})|\Upsilon^{\alpha\beta}\rangle
\langle\Upsilon^{\gamma\delta}|(U^{00} \otimes R^{\mu\nu\dagger}U^{\mu\nu})|\Upsilon^{00}\rangle.
\end{eqnarray}
The last equality in Eq.(30) follows from  ${\rm tr}[U^{\alpha\beta\dagger}U^{\gamma\delta}] = 4\delta_{\alpha\gamma}\delta_{\beta\delta}$ and
\begin{eqnarray}
{\rm tr}[R^{\mu\nu}U^{\alpha\beta\dagger}U^{\mu\nu\dagger}] & = & 
4\langle\Upsilon^{00}|(U^{00} \otimes U^{\mu\nu\dagger}R^{\mu\nu})|\Upsilon^{\alpha\beta}\rangle, \nonumber \\
{\rm tr}[U^{\mu\nu}U^{\gamma\delta}R^{\mu\nu\dagger}] & = & 
4\langle\Upsilon^{\gamma\delta}|(U^{00} \otimes R^{\mu\nu\dagger}U^{\mu\nu}|\Upsilon^{00}\rangle.
\end{eqnarray}
Substituting Eq.(30) into Eq.(29), we obtain Eq.(15).  This completes what we set out to do.

Interestingly, from Eq.(23), we note that ${\cal N}[\Lambda^{\Xi, {\cal E}_0}_{B_1B_2}(|\Psi\rangle_{B_1B_2}\langle\Psi|)] = \max\{0,\ \sin2\epsilon\}$ when $q = 0$ or while ${\cal G}[\Xi] = 1/4$.  That is, the Smolin state enables the teleportation of all the entanglement associated with the input state $|\Psi\rangle_{A_1A_2}$ albeit with $\langle\Psi|\Lambda^{\Xi, {\cal E}_0}(|\Psi\rangle\langle\Psi|)|\Psi\rangle = (1 + \sin^22\epsilon)/2$, which approaches 1 as $\epsilon \rightarrow \pi/4$.  This may seem to be an obvious counterexample to our above conjecture.  However, we wish to point out that the entanglement associated with the Smolin state is actually between any sinlge qubit and the remaining three qubits, since the negativity say between $A_3$ and $A_4B_1B_2$ is ${\cal N}[\Xi^{\rm S}_{A_3(A_4B_1B_2)}] = 1$; and any one particle loss results in the completely random state.  ${\cal G}[\Xi] = 1/4$ is in agreement with the fact that there is zero entanglement between $A_3A_4$ and $B_1B_2$ in the Smolin state.  Hence, ${\cal G}[\chi]$ really describes the entanglement between $A_3A_4$ and $B_1B_2$, and there is no contradiction.  Multipartite entanglement is indeed much more interesting.

Lastly, let us look at $\Xi^{\rm GHZ} \equiv |\Psi^0_{\rm GHZ}\rangle\langle\Psi^0_{\rm GHZ}|$ and $\Xi^{\rm W} \equiv |\Psi^1_{\rm W}\rangle\langle\Psi^1_{\rm W}|$, where $|\Psi^0_{GHZ}\rangle \equiv (|0000\rangle + |1111\rangle)/\sqrt{2}$ and $|\Psi^1_{\rm W}\rangle = (u^1 \otimes u^0 \otimes U^{00})|\Psi^0_{\rm W}\rangle$, $|\Psi^0_{\rm W}\rangle \equiv (|0001\rangle + |0010\rangle + |0100\rangle + |1000\rangle)/2$.  These give
\begin{eqnarray}
\langle\Upsilon^{00}|\Xi^{\rm GHZ}|\Upsilon^{00}\rangle & = & \frac{1}{2}\cos^2\theta_{12}, \nonumber \\
\langle\Upsilon^{00}|\Xi^{\rm W}|\Upsilon^{00}\rangle & = & \frac{1}{16}[2 + \sin2\theta_{12} + 2\sin(\theta_{12} + \phi_{12}) + 2\cos(\theta_{12} - \phi_{12}) + \sin2\phi_{12}],
\end{eqnarray}
and ${\cal G}[\Xi^{\rm GHZ}] = 1/2$ when $\theta_{12} = 0$; while ${\cal G}[\Xi^{\rm W}] = 1/2$ when $\theta_{12} = \phi_{12} = \pi/4$.  For $|\Psi\rangle_{A_1A_2}$, the negativities of the teleported state are respectively,
\begin{eqnarray}
{\cal N}[\Lambda^{\Xi^{\rm GHZ}, {\cal E}_0}_{B_1B_2}(|\Psi\rangle_{B_1B_2}\langle\Psi|)] & = & 
\max\{0,\ \cos2\theta_{12}\sin2\epsilon\}, \nonumber \\
{\cal N}[\Lambda^{\Xi^{\rm W}, {\cal E}_0}_{B_1B_2}(|\Psi\rangle_{B_1B_2}\langle\Psi|)] & = & 
\max\{0,\ \frac{1}{2}\sin2\phi_{12}\sin2\epsilon\}.
\end{eqnarray}
First, we note that with $\theta_{12} = 0$, the first equation in Eq.(34) is a well-known result.  That is, the four-qubit GHZ state enables the faithful teleportation of a partially known two-qubit state if Alice performs projective measurement in the basis consisting of tensor products of Bell states.  Second, if one considers only $\theta_{12} = \phi_{12} = 0$ for $\Xi^{\rm W}$, then $\langle\Upsilon^{00}|\Xi^{\rm W}|\Upsilon^{00}\rangle = 1/4$ and ${\cal N}[\Lambda^{\Xi^{\rm W}, {\cal E}_0}_{B_1B_2}(|\Psi\rangle_{B_1B_2}\langle\Psi|)] = 0$.  This is an equally famous null result, if Alice is confined only to projective measurement of the above sort.  Together with the above examples, it clearly illustrates that in studying the feasibility of a given four-qubit entangled state for entanglement teleportation, it is necessary to analyze the generalized singlet fraction, Eq.(16).

In conclusion, we have shown that the teleportation protocol ${\cal E}_0$, when used with an arbitrary four-qubit mixed state $\Xi$ as a resource, acts as a generalized depolarizing bichannel with probabilities given by the maximally entangled components of the resource, Eq.(14).  We define the generalized singlet fraction ${\cal G}[\Xi]$, Eq.(16), which is a necessary quantity to analyze in order to determine the usefulness of $\Xi$ as a resource for ${\cal E}_0$.  We could analogously define the maximal generalized singlet fraction
\begin{equation}
{\cal G}_{\max}[\Xi] \equiv \max_{\theta_{12}, \phi_{12}, U}
\langle\Upsilon^{00}(\theta_{12}, \phi_{12})|(U^{00} \otimes U)\Xi(U^{00} \otimes U)|\Upsilon^{00}(\theta_{12}, \phi_{12})\rangle,
\end{equation}
where we have in addition the maximization over the group of all unitary transformations $U$ on ${\cal C}^2 \otimes {\cal C}^2$.  Eq.(15) together with the compactness of this group guarantees that ${\cal G}_{\max}[\Xi]$ and hence the maximal teleportation fidelity
\begin{equation}
\Phi[\Lambda^{\Xi, {\cal E}_{\rm opt}}_{B_1B_2}] = \frac{1}{5} + \frac{4}{5}{\cal G}_{\max}[\Xi]
\end{equation}
is achievable via $R^{\mu\nu} = R^{\mu\nu}_{\rm opt}$.  This raises the following interesting problem, among many others, in addition to the proof of our conjecture.  While it is expected that there will still be states with some critical negativity ${\cal N}'_{\rm crit}$, which after being teleported with nonclassical fidelity are however left with zero entanglement; the question is if ${\cal N}'_{\rm crit} < {\cal N}_{\rm crit}$?  It is our hope that results presented and issues raised in this paper would contribute to our understanding of multipartite mixed state entanglement.

\end{document}